\documentclass{amsart}
\usepackage{graphicx}
\usepackage{epsf}
\theoremstyle{definition}

\theoremstyle{remark}

\numberwithin{equation}{section}

\newcommand{\bt}{\begin{tabular}}
\newcommand{\et}{\end{tabular}}
\newcommand{\bi}{\begin{itemize}}
\newcommand{\ei}{\end{itemize}}
\newcommand{\bbm}{\begin{minipage}[h]{6cm} ${}$ \\}
\newcommand{\eem}{\\ ${}$ \end{minipage}}
\newcommand{\bmm}{\begin{minipage}[h]{1.2truecm} ${}$ \\}

\newcommand{\be}{\begin{equation}}
\newcommand{\ee}{\end{equation}}

\begin{document}


\title[Majorana and the Thomas-Fermi model: a comment to
physics/0511222]{Again on Majorana and the Thomas-Fermi model: a comment to
physics/0511222}%
\author{S. Esposito}%
\address{{\it S. Esposito}: Dipartimento di Scienze Fisiche,
Universit\`a di Napoli ``Federico II'' \& I.N.F.N. Sezione di
Napoli, Complesso Universitario di M. S. Angelo, Via Cinthia,
80126 Napoli ({\rm Salvatore.Esposito@na.infn.it})}%


\begin{abstract}
We comment on a recent paper announcing the discovery of a previously unknown publication
of Ettore Majorana on the Thomas-Fermi atomic model. In pointing out that such a
publication was not written by Majorana, we correct some misunderstandings and clarify
the historical and scientific relevance of the ``forgotten publication".
\end{abstract}

\maketitle

\noindent In a recent paper \cite{GR} the authors claim to have discovered ``a forgotten
publication of Ettore Majorana on the improvement of the Thomas-Fermi statistical model",
quoting also the corresponding reference to the Italian journal ``Il Nuovo Cimento"
\cite{EM}. Such a publication refers to a communication at the XXII General Meeting of
the Italian Physical Society, made by Majorana on December 29, 1928, where he reported on
some peculiar applications to atomic and molecular spectra of the statistical model
introduced by Thomas \cite{Thomas} and, independently, by Fermi \cite{FNM43} few years
earlier. This paper has served to give a (partial) re-analysis of the role played by
Majorana in the formulation of the Thomas-Fermi atomic model. We realize the relevance of
this publication \cite{GR} and its integration among the few public appearances of Ettore
Majorana in the scientific scenario of his epoch. However, in view of the potential
importance of this subject, we have to remark that some of the conclusions derived in it
should be critically revised.

The reason for considering the communication in \cite{EM} as a ``forgotten publication"
is based on the fact that it is not reported in the list of publications by Ettore
Majorana, first given by a friend and colleague of him, Edoardo Amaldi \cite{Amaldi}, and
subsequently adopted by other authors. This occurrence, together with the fact that the
Majorana communication did not receive any mention in the works by Fermi and his
associates (Amaldi was one of them), has led to postulate that ``the Majorana proposal
was not accepted by Enrico Fermi for years, and forgotten", while ``at the end, Fermi
eventually fully accepted Majorana improved scheme, and exploited it in the conclusive
paper on the subject" (this paper is the one in Ref. \cite{FNM82}).

However, such a suggestive conclusion appears, at a deeper analysis, slightly incorrect
and a much simpler explanation seems to emerge.

First of all, a careful reading of the paper in \cite{EM} reveals that it was {\it not}
written by Majorana, although the corresponding talk at the Meeting was instead delivered
by him, as appears clearly from the accurate study of his style through his published and
unpublished notes, an account of which can be found in Ref. \cite{Volumetti}. However,
even without invoking such an expertise, it is a matter of fact that the material
published in Ref. \cite{EM} in ``Il Nuovo Cimento", which was the official organ of the
Italian Physical Society, referred to the formal minutes of the General Meeting of that
Society, and even the scientific communications (with the sole, probable, exception of a
couple of them) were written by the secretary of the Society, G. Dalla Noce (a former
engineer and, then, a theoretical physicists).

In this respect the paper in \cite{EM} cannot be regarded as a publication by Ettore
Majorana, and thus it was never mentioned as such by Amaldi, Fermi and others.

Nevertheless, useful information may be deduced by what contained in \cite{EM}, the most
intriguing one being that the young Majorana was invited to deliver that talk well {\it
before} to have got his ``laurea" degree in Physics. In fact, he graduated in July 1929,
and the topics covered in his communication were completely different from those
discussed in his master thesis (on the theory of radioactive nuclei) and even only
marginally related to his first article \cite{first} on Atomic Physics, in collaboration
with G. Gentile Jr, also published before his graduation in Physics. Evidently, the
relevance of the work performed by Majorana at the Institute of Physics in Rome,
``inspired" by discussions with Fermi and coworkers, was well recognized by Fermi himself
who, as in some other occasions, urged the young researcher to make publicly available
the conclusions obtained.

The attitude of Majorana not to spread the results of a given research, until its perfect
refinement (according to his hypercritical judgement) was completed or when they were
considered premature, manifested here in the fact that he decided to not publish them in
the form of a regular article. This is clearly stated at the end of the communication in
\cite{EM}, ``the researches performed till now are still too much scarce to fully
appreciate such results", and, as said, it is typical of the personality of Majorana.
Likely, the fact that the Meeting was held in the Institute of Physics, where he was
currently performing his studies, played some role in Majorana's decision to give a talk.
\\
In any case it is quite evident that Fermi was well aware of the work done by Majorana on
the statistical model of atoms, and pushed his student to present it in the occasion of
the General Meeting of the Italian Physical Society. Note also that the communication by
Majorana was ``sandwiched" between two other ones by Fermi \cite{EM}, and no question was
posed by the conveners (including Fermi).

An accurate historical reconstruction of the appearance and first developments of the
statistical model of atoms, which focuses on the main results achieved by the group of
Fermi in Rome and on the role played by Majorana, can be found in Ref. \cite{DiGrezia}.
Here we do not repeat what reported there, but just comment on some scientific, rather
than historical, issues. The systematic contributions by Majorana may be found in Ref.
\cite{Volumetti}, where five notebooks by Majorana are translated into English and
published for the first time. The original material behind the communication in \cite{EM}
is there reported at pages 116-117.

According to Majorana, the topic studied is the {\it second approximation for the
potential inside the atom}, with a generalization (rather than an improvement) of the
Thomas-Fermi model of neutral atoms of atomic number $Z$ to those ionized $n$ times
(including the case $n=0$). The starting point is the physical fact that such a potential
is defined {\it up to an additive constant} $C$, as already clearly stated by Fermi as
long as in his 1927 paper in Ref. \cite{FNM43}. This occurrence is exploited by Majorana
in order to shift the attention from the local potential $V_0$ in a given point inside
the atom or the ion (that is the potential of the nucleus and $Z-n$ electrons), to the
effective potential $V$ acting on one electron in that point (that is the potential of
the nucleus and $Z-n-1$ electrons). The two potentials are connected, approximatively, by
a simple scaling relation that Majorana writes as %
\be \label{1} %
\nabla^2 V = \frac{Z-n-1}{Z-n} \, \nabla^2 V_0
\ee %
(instead Fermi, in 1934 \cite{FNM82}, will write as
\be \label{2} %
V = \frac{Z-n-1}{Z-n} \, V_0 ,
\ee %
that is the scaling relation is assumed here to hold directly on the potentials). Since
$V_0$ satisfies the Poisson equation with a charge density $\rho$, $\nabla^2 V_0 = - 4
\pi \rho$, the effective potential $V$ is that generated by a rescaled charge density
$\rho (Z-n-1)/(Z-n)$, thus taking into account the finite charge of the given electron on
which $V$ (rather than $V_0$) acts. Although in a completely different context, the
procedure is rather similar to that now adopted in the renormalization of physical
quantities in modern gauge theories. 
\\
The effective potential is
\be \label{3}%
V = \frac{Z e}{r} \, \varphi \left( \frac{r}{\mu} \right) \, + \, C ,
\ee %
where $\varphi$ is the Thomas-Fermi screening function ($\mu$ is a dimensionful
parameter) and the constant $C$ is interpreted as value of the potential at the boundary
of the atom (or the ion), which thus acquires a finite radius $r_0$ ($=\mu x_0$):
\be \label{4}%
C = \frac{(n+1)e}{\mu x_0} .
\ee %
Consequently, the maximum energy $U$ (which, in general, is negative) of one bounded
electron is different from zero and proportional to $C$: %
\be \label{5} %
U = - C e .
\ee %
From this expression, the Rydberg corrections to the energy levels of a given atom or ion
may be easily deduced.
\\
As appears very clearly from Ref. \cite{EM} and what expounded above, the key role in the
Majorana approach is played by the additive constant $C$, rather than by the finite
radius $r_0$. Although the question of the effective atomic radius, as emphasized in
\cite{GR}, rests on solid physical grounds (an atom or an ion has indeed a finite
extension) and was referred to very early, it is not very relevant for practical
applications at a given level of approximation. In a sense, as also shown by Majorana, a
finite radius $r_0$ can always be obtained by requiring a finite range for the potential
$V$, and this is simply achieved through the introduction of a suitable constant $C$,
which can always be added to the potential. In this respect, it seems not correct to
interpret \cite{GR} the term $C$ as a Lagrange multiplier which constraints the atom to
have a finite radius, and, in any case, this does not correspond to what done by
Majorana.

Furthermore, regarding the possible refusal or acceptance of the Majorana viewpoint by
Fermi, the following points should be taken into account.
\\
In almost all the papers by Fermi on this subject, where he reports also few theoretical
calculations about the Thomas-Fermi model, the question of the definition of the
potential inside the atom {\it up to an additive constant} is always explicitly
asserted\footnote{In particular note that the correct expression for the electron density
$n$ is that in Eq. (5) of Ref. \cite{FNM43} (or similar equations in Refs. \cite{FNM63},
\cite{FNM82}), with the potential denoted by Fermi with $v$, and not that reported in Eq.
(1) of Ref. \cite{GR}, with the potential denoted with $V$: these two potentials differ,
exactly, for an additive constant!} \cite{FNM43}, \cite{FNM63}, \cite{FNM82}.
\\
In the papers dealing with neutral atoms (starting from Ref. \cite{FNM43} of 1927),
though not expressly mentioned, the constant is {\it chosen} to be zero or, in other
words, it is imposed for simplicity that the atomic radius be infinite (the upper limit
in the integral in Eq. (10) of Ref. \cite{FNM43} is $\infty$).
\\
In the article in Ref. \cite{FNM63} of 1930-1, discussing the energy spectra of ionized
atoms, a finite radius $r_0$ for them has to be explicitly assumed, as well known, but
this is introduced along the same lines followed by Majorana, that is the additive
constant playing a key role (see the discussion in \cite{FNM63} from Eq. (1) to Eq. (5)).
However it is evident as well that the subtleties envisaged by Majorana, in
distinguishing the local potential from the effective one, are not present at all (they
can be neglected at the degree of approximation considered there). This is easily
recognized by noting that the expressions quoted by Majorana in \cite{EM} (or, better, in
Sect. 15 of Volumetto 2 in \cite{Volumetti}) reduce to those reported by Fermi in
\cite{FNM63} by replacing $n+1$ with $n$ (or $z$, as Fermi originally denoted the
ionization degree).
\\
Finally, in the last paper of the Fermi group on the applications of the Thomas-Fermi
model, published in 1934 \cite{FNM82}, the ``choice" (according to Fermi and Amaldi
wording) of the atomic potential is made by following completely\footnote{With the
exception mentioned above in Eq. (\ref{2}).} the reasoning of Majorana, as pointed out in
Ref. \cite{GR}. This is {\it  required} by the improvement in the degree of approximation
pursued here by the authors, with respect to previous calculations, which now include
relativistic corrections too. In fact, the ``second approximation" (as termed by
Majorana) for the effective potential depends on the ratio between the degree of
ionization and the atomic number (remember that such ratio is written, approximatively,
as $n/Z$ in \cite{FNM63} and, correctly, as $(n+1)/Z$ in \cite{FNM82} and \cite{EM}), so
that it has some (small) influence only for lighter elements, as considered only in Ref.
\cite{FNM82}. Here, in fact, $Z \geq 10$, while in Ref. \cite{FNM63} the calculations
were carried out for $Z=23, 50, 82$ (in previous calculations, the degree of
approximation is rough and then the effect considered here is irrelevant).

From what discussed above, assuming that Fermi was aware of the Majorana approach (see
the acknowledgment in Ref. \cite{EM}), it seems rather reasonable that Fermi did not
refuse that approach, but simply used it only when explicitly necessary, according to his
general attitude to avoid unnecessary mathematical complications. The same applies to the
Fermi and Segr\`e paper \cite{FS} on hyperfine structures of atomic spectra of 1933; in
particular, note the discussion from Eq. (13) to Eq. (17) there.
\\
In practice, what can be safely deduced from the appearance of Ref. \cite{EM} is the
following. Majorana discussed with Fermi (and, maybe, others) a refinement of the
statistical model of atoms and some applications of it (on the chemical bonds and the
Roentgen spectra), and Fermi, convinced of the theoretical relevance of the results
obtained by Majorana, urged him to communicate those results at the General Meeting of
the Italian Physical Society in 1928, in between two talks of his since Majorana was
still not graduated. Majorana, on the other hand, considering premature his own results
according to his hypercritical feeling, decided to not publish them in a regular article.
Fermi applied, even tough in a slightly different way (see above), the refinement of the
statistical model only when required by the accuracy of the results to be obtained, and
the lacking\footnote{Instead, in many other occasions, Fermi was always very diligent to
mention the contributions by Majorana or other associates. Just as an example related to
the topic considered here, see the reference to the Gentile and Majorana work
\cite{first} in his papers.} of the acknowledgment to Majorana's work is probably a too
weak argument to rediscuss the relations between the two scientists.

In conclusion, while the important role played by Majorana, in the formulation and first
developments of the Thomas-Fermi statistical model of atoms, is undoubtedly recognized,
as discussed extensively in \cite{DiGrezia}, the relevant theoretical issue behind the
work related to Ref. \cite{EM} appears to be what we can improperly term the
renormalization procedure of the Thomas-Fermi potential. However, it seems quite
reductive to focus only on this contribution, while Majorana introduced some other
generalizations of the model, disregarding his solution of the Thomas-Fermi non-linear
equation \cite{AJP}, \cite{IJTP}. Almost all on this unpublished (by the author) work can
be found in the book in \cite{Volumetti}, and the interested reader can usefully follow
the related discussion in Ref. \cite{DiGrezia}.

We end with the hope, also envisaged in \cite{GR}, that future recognitions of the
outstanding but unknown work performed by Majorana in completely different areas of
Physics, shall ``not rely on more or less arbitrary reconstructions from fragmentary
unpublished sources", but take benefit of already established results.

We warmly thank Ettore Majorana Jr for having pointed out the appearance of the preprint
in \cite{GR} and for valuable discussions.


\end{document}